# Superconductivity and thermal properties of sulphur doped FeTe with effect of oxygen post annealing

V. P. S. Awana[*], Anand Pal, Arpita Vajpayee, Bhasker Gahtori and H. Kishan

*National Physical Laboratory, Dr. K.S. Krishnan Marg, New Delhi 110012, India*

Abstract

Here, we report the synthesis and characterization of sulphur-substituted iron telluride i.e. FeTe$_{1-x}$S$_x$; (x = 0-30 %) system and study the impact of low temperature oxygen (O$_2$) annealing as well. Rietveld analysis of room temperature x-ray diffraction (XRD) patterns shows that all the compounds are crystallized in a tetragonal structure (space group *P*4/*nmm*) and no secondary phases are observed. Lattice constants are decreased with increasing S concentration. The parent compound of the system i.e. FeTe does not exhibit superconductivity but shows an anomaly in the resistivity measurement at around 78 K, which corresponds to a structural phase transition. Heat capacity C$_p$(*T*) measurement also confirms the structural phase transition of FeTe compound. Superconductivity appears by S substitution; the onset of superconducting transition temperature is about 8 K for FeTe$_{0.75}$S$_{0.25}$ sample. Thermoelectric power measurements *S*(*T*) also shows the superconducting transition at around 7 K for FeTe$_{0.75}$S$_{0.25}$ sample. The upper critical fields $H_{c2}$(10%), $H_{c2}$(50%) and $H_{c2}$(90%) are estimated to be 400, 650 and 900 kOe respectively at 0 K by applying Ginzburg Landau (GL) equation. Interestingly, superconducting volume fraction is increased with low temperature (200 $^o$C) O$_2$ annealing at normal pressure. Detailed investigations related to structural (*XRD*), transport [*S*(*T*), *R*(*T*)*H*], magnetization (AC and DC susceptibility) and thermal [*C*$_p$(*T*)] measurements for FeTe$_{1-x}$S:O$_2$ system are presented and discussed.

Key Words: FeTe$_{1-x}$S$_x$, Structure, Magnetization Superconductivity, Thermo-power.

PACS: 74.25.-q, 74.62.Bf, 74.62.Dh

[*]Corresponding Author
Dr. V.P.S. Awana
Fax No. 0091-11-45609310: Phone No. 0091-11-45608329
e-mail-awana@mail.nplindia.ernet.in;
www.freewebs.com/vpsawana/

Introduction

The discovery of the iron-based layered superconductor LaFeAs($O_{1-x}F_x$) having critical temperature ($T_c$) of 26 K reported by Kamihara et al. had a great impact to researchers in condensed-matter physics [1]. The $T_c$ was raised by applying pressure ($T_c^{onset}$ = 43 K) [2] or substitution of smaller rare earth ion for the La site ($T_c$ = 55 K for SmFeAsO$_{1-x}$F$_x$) [3]. Soon after these discoveries, researchers reported three other crystal structure types of layered Fe compounds to be superconducting, including Ba$_{1-x}$K$_x$Fe$_2$As$_2$ [4], Li$_{1-x}$FeAs [5] and FeSe [6]. These materials share common structural square layers with Fe coordinated with either a pnictogen or a chalcogen. After the discovery of superconductivity in FeSe, it was natural to ask whether chemical substitutions, either to the Se-site or the Fe-site, have any effects on $T_c$. The $T_c$ of FeSe is increased up to 15 K by partial substitution of Te or S for Se and it is also increased up to ~27 K by applying hydrostatic pressure of 1.48 GPa [7-10]. Density functional calculations for FeS, FeSe and FeTe indicated the strength of spin density wave (SDW) in FeTe and the possibility higher $T_c$ in doped FeTe or Fe(Se,Te) alloy than FeSe [11]. Tetragonal FeSe is a superconductor with a transition temperature $T_c$ of 8 K and shows enhancement of $T_c$ with applying pressure. Tetragonal FeTe is structurally analogous to superconducting FeSe, but does not show superconducting transition and rather it undergoes a structural phase transition from tetragonal to orthorhombic at around 80 K [8, 12]. It is observed that the phase transition depends upon the excess-Fe concentration into the compound. If the excess-Fe concentration are low the structural transition would be tetragonal to monoclinic [13,14]. No sign of superconductivity was found in FeTe system after applying the pressure of 1.6 GPa, however structural phase transition temperature shifted to lower temperature [15]. Okada et al. demonstrated that unlike in FeAs-based compounds, no superconductivity was detected at high pressures of even up to 19 GPa in FeTe$_{0.92}$, although the 80K anomaly at atmospheric pressure was suppressed by applying pressure [16]. In any case, it appeared that chemical pressure in Fe-Te layer might still induce superconductivity in this system. Mizuguchi et al. used the same idea and observed the superconductivity at about 10 K in sulphur (S) substituted FeTe [17]. Further, with exposing the FeTe$_{0.8}$S$_{0.2}$ sample to the humid air, the zero resistivity temperature $T_c^{zero}$ and the superconducting volume fraction were enhanced up to 7.2 K and 48.5 %, respectively [18]. The process took several days of humidity exposure [18]. Recently same authors also reported the evolution of superconductivity by oxygen annealing for several hours at low temperatures [19].

We report here the synthesis and induction of superconductivity by sulphur (S) substitution in $FeTe_{1-x}S_x$ system. We obtained superconductivity with $T_c$ of up to 8 K for 10-30% doping levels. Since S has a smaller ionic radius than Te, the Te site S substitution creates chemical pressure in the unit cell and the superconductivity appears. The upper critical field of the $FeTe_{1-x}S_x$ is estimated of up to 900 kOe. Further, the superconducting volume fraction is increased with low temperature $O_2$ annealing. Interestingly, the present situation is different from that reported in ref. 18, our samples are not exposed to air for several days, but we annealed the samples in $O_2$ atmosphere at 200 °C only for 12 hours as suggested by Mizuguchi et al. [19] and comparable results are obtained.

## Experimental

Polycrystalline samples of $FeTe_{1-x}S_x$; (x = 0-30%) were synthesized by the solid-state reaction route. The stoichiometric ratio of high purity (>3N) Fe, Te and S were ground, palletized and encapsulated in an evacuated quartz tube. The encapsulated tube was then heated at 750 °C for over 24 hours and slowly cooled to room temperature. Sintered pellets were again ground, pelletized in a rectangular shape, sealed in an evacuated quartz tube and re-sintered at 750 °C for 12 hours. As synthesized $FeTe_{1-x}S_x$ samples were further annealed in flow of $O_2$ at 200 °C for 12 hours and cooled to room temperature. The X-ray diffraction patterns of the samples were obtained with the help of a Rigaku diffractometer using $CuK_\alpha$ radiation. All physical property measurements including magneto-transport $R(T)H$, thermoelectric power $S(T)$, heat capacity $C_p(T)H$ and magnetization (AC and DC) were carried out using *Quantum Design* PPMS (Physical Property Measurement System).

## Results and Discussion

The room temperature X-ray diffraction (XRD) pattern for $FeTe_{1-x}S_x$; x=0.0, 0.10, 0.25 & 0.30 samples along with their Rietveld analysis are presented in Figure l(a). The structure of $FeTe_{1-x}S_x$ for all the compositions of x was refined with the tetragonal space group *P*4/*nmm* (space-group no.129). Fe is located at coordinate position (¾, ¼, 0) and Te/S at (¼, ¼, z) in $FeTe_{1-x}S_x$ compounds. All the compounds are crystallized in a tetragonal structure and no secondary phases are observed. The lattice constants, volume of unit cells, 'z' coordinates of Te/S and Rietveld refinement parameters are listed in Table 1. Both lattice parameters '*a*' and '*c*' decreased monotonically with increasing nominal S content due to the substitution of S at Te site. The lattice parameter '*a*' and '*c*' are 3.821(9) and 6.285(5) Å for FeTe and the same are reduced to 3.805(7) and 6.224(2) Å for $FeTe_{0.70}S_{0.30}$ compound respectively. Variation of lattice parameters and cell volume (V) with increasing S content is shown in Figure 1(b).

Figure 2 represents the temperature dependence of resistance for FeTe and $FeTe_{0.75}S_{0.25}$. For FeTe no sign of superconductivity is observed but an anomaly is observed at around 78 K, which corresponds to a structural phase transition (marked by arrow in Figure 2) [8,9,12,13,18]. On the other hand superconductivity is clearly observed in S substituted FeTe samples at low temperature below 10 K. More specifically, the $T_c^{onset}$ and $T_c^{R \rightarrow 0}$ are at ~10.0 and 5.5 K respectively for $FeTe_{0.75}S_{0.25}$ sample. The 78 K anomaly in resistivity measurements of FeTe also exhibits sufficient hysteresis of up to 6-10 K in cooling and warming cycles of *R*(*T*) measurements. One can presume that the *R*(*T*) anomaly occurs close to say 70 K in FeTe. In order to study the nature of 78 K anomaly for FeTe, we further performed specific heat measurement under magnetic fields of 0, 7 and 14 Tesla, and the results are shown in inset of Figure 2. The absolute value of $C_P$ at 200 K is around 57 J/mol-K. As the temperature is decreased further the $C_P$ goes down continuously. A distinct hump/kink is seen in $C_P(T)$ at around 69 K for zero field. Interestingly this is nearly the same temperature, where the metallic step is observed in *R*(*T*) at zero-field.  Further, it is found that neither hump position nor magnitude of $C_p$ changes in presence of applied magnetic fields of up to 14 Tesla. This means that after applying the magnetic field of as high as 14 Tesla the $C_P$ hump is yet observed at the same temperature having the same magnitude. It is more likely that the $C_P(T)$ anomaly at around 70 K is mainly due to the reported structural phase transition and possible contribution of any magnetic state such as SDW is very less. Further, the structural phase transformation is quite robust, as the same is not altered even slightly under as high as 14 Tesla applied field.

The temperature dependence of thermoelectric power $S(T)$ for superconducting FeTe$_{0.75}$S$_{0.25}$ sample is shown in Figure 3. The absolute value of $S$ is negative, which indicates towards the electron type conductivity in this system. The room temperature thermoelectric power $S^{300K}$ is around 16 µV/K. Superconducting transition ($T_c$) is seen as $S$=0 at ~7 K, which confirms the occurrence of superconductivity in this sample. A small upside hump is also seen before onset of superconductivity. The $S(T)$ results of FeTe$_{0.75}$S$_{0.25}$ sample clearly shows that the compound is superconducting below 7 K and has electrons as majority charge carriers.

The AC susceptibility, $\chi = \chi' + \chi''$, in the absence of DC magnetic field is measured in a polycrystalline sample of FeTe$_{0.75}$S$_{0.25}$ to probe the details of magnetic and superconducting transitions. Figure 4 represents the plot of real ($M'$) and imaginary ($M''$) components of AC susceptibility as a function of temperature obtained at the frequencies 33, 333, 999, 3333, 6666 & 9999 Hz at AC drive field amplitude of 10 Oe. Real part of AC magnetization ($M'$) shows the superconducting onset temperature at ~7 K for all the frequencies, which is same as being observed in $R(T)$ and $S(T)$ measurement respectively in Figures 2 and 3 for the same sample. The $M'(T)$ curves do not show the typical two-step transition due to coupling (Inter-grain, at low $T$) and intrinsic (intra-grain, near $T_c$) response [20-22]. Only a well-defined coupling (inter granular) peak is observed for all the frequencies in the imaginary component of susceptibility ($M''$) at a characteristic temperature $T_p$. This peak arises when the AC field penetrates through the intergranular region just to the center of the sample. On increasing the frequency, the peak intensity increases a little but almost there is no change in peak temperature ($T_p$).

The temperature variation of the AC susceptibility measured at different AC drive field amplitudes 0.1, 0.5, 1, 2 & 4 Oe of FeTe$_{0.75}$S$_{0.25}$ sample is shown in Figure 5. The diamagnetic onset temperature is ~ 7 K for all the fields i.e. it is constant at different fields. It can be noticed that here also the AC susceptibility data show only the coupling (inter-granular) peak i.e. intrinsic (intra-granular) peak is not present in the figure for all the values of AC field amplitudes. The enlarged view of $M''$ data at 0.1 Oe is shown in the inset of Figure 5. In the imaginary ($M''$) components of AC susceptibility the inter-granular peak temperature $T_p$ decreased from 5.4 K to 2.5 K, when the applied field amplitude is increased from 0.1 Oe to 2 Oe. It is also noted that values of maximum $M''(T_p)$ is increasing monotonically with increasing field.

The absence of intrinsic (intra-granular) peak at near $T_c$ region suggests that the grains are well diffused in the sample even at the frequency of 9999 Hz and the field of 4 Oe. This implies that coupling between the grains is strong enough and de-coupling of grains in the polycrystalline $FeTe_{0.75}S_{0.25}$ sample does not take place at this magnitude of AC field and of frequency. This is unlike the HTSc and oxy-pnictides [20-24].

Though the bulk superconductivity is confirmed by onset of diamagnetism in real part of AC susceptibility ($M'$) but the signal is weak and apparently broad as well. This behavior is an indication of low volume fraction weak superconductivity of the studied system. The superconducting volume fraction of the studied $FeTe_{1-x}S_x$ system could be increased after annealing the same samples in $O_2$ flow at 200 $^oC$; the results will be presented and discussed in next sections.

In order to determine the upper critical field of the $FeTe_{0.75}S_{0.25}$ sample $R(T)$ curves are measured under different magnetic fields of up to 110 kOe. The superconducting transition zone of $R(T)H$ measurements is shown Figure 6. The upper critical field is determined using different criterion of $H_{c2}=H$ at which $\rho=90\%\rho_N$ or $50\%\rho_N$ or $10\%\rho_N$ where $\rho_N$ is the normal resistance or resistivity at about 10 K. The $H_{c2}$ variation with temperature is shown in upper inset of Figure 6. To determine $H_{c2}(0)$ value, we applied Ginzburg landau (GL) theory. The GL equation is:

$$H_{c2}(T)= H_{c2}(0)*(1-t^2)/(1+t^2)$$

Where $t=T/T_c$ is the reduced temperature [25]. The fitting of experimental data is done according to the above equation, which not only determines the $H_{c2}$ value at zero Kelvin [$H_{c2}(0)$] but also determines the temperature dependence of critical field for the whole temperature range. $H_{c2}(10\%)$, $H_{c2}(50\%)$ and $H_{c2}(90\%)$ are estimated to be 400, 650 and 900 kOe respectively at 0 K. The results are plotted in inset of Figure 6. Mizuguchi et al. [17] also estimated the upper critical field for $FeTe_{0.8}S_{0.2}$ at 0 K [$H_{c2}(0)$] not by applying the GL equation but from the linear extrapolation of $H_c(T)$ plots. They estimated values of [$H_{c2}(0)$] i.e. ~ $H_{c2}$(onset), $H_{c2}$(midpoint) and $H_{c2}$(zero) of 102, 78, 56 Tesla in comparison to present case.

It is interesting that though the AC magnetization results (Figure 4 & 5) indicate towards the weak superconductivity in terms of low volume fraction, the $R(T)H$ measurements (Figure 6) yet give reasonably high $H_{c2}(0)$. To overcome the low superconducting volume fraction situation, we followed the low temperature $O_2$ annealing route as suggested by Mizuguchi et al. [19]. The results of both DC and AC susceptibilities for these $O_2$ annealed $FeTe_{1-x}S_x$:$O_2$ samples are presented in Figures 7(a) and 7(b) respectively. Superconductivity is observed in both FC (field-cooled) and ZFC (zero-field-cooled) situations at transition temperature ($T_c$) around 8.5 K. Interestingly, $T_c$ is nearly same for various $FeTe_{1-x}S_x$:$O_2$ samples with x ranging from 0.10 to 0.30. Te site S substitution provides the chemical pressure to unit cell and hence the appearance of superconductivity. It is seen from lattice parameters variation (Fig. 1(b)) that S monotonically substitutes at Te site till x =0.30 and no un-reacted lines are seen in XRD (Fig.1(a)). May it be that $T_c$ saturates around 10 K with application of chemical pressure beyond x =0.10. This is true for the results reported by Mizuguchi et al. as well; they found same $T_c$ in magnetization measurements for $FeTe_{0.9}S_{0.1}$ and $FeTe_{0.8}S_{0.2}$ samples [17]. In the absence of the superconductivity in FeTe system under possible applied physical pressures [15,16], it is difficult to cross check if $T_c$ of this system could saturate above a certain chemical pressure. For now what appears is that $T_c$ of $FeTe_{1-x}S_x$ remains nearly invariant at close to 10 K for x = 0.10 to say 0.30 [ref. 18 and present study]. The only difference is to be noted that with increasing value of x though $T_c$ is nearly same, but the superconducting volume fraction is seemingly maximum for x = 0.30 sample and the same is least at both ends, i.e., for x = 0.10.. This is inter comparative estimate only as adjudged from the ZFC signal, which consists of both the superconducting (Meissner) and shielding (non dissipating current on surface) contributions. Infact the FC signal (Meissner only contribution) is quite small and overlapping with each other near base line and hence we compared the ZFC signal. The exact estimation of superconducting volume fraction in a superconductor is tricky due to presence of pinning centers etc. within the intrinsic superconductor. At this juncture we only conclude that comparatively the x=0.30 sample possesses higher superconducting volume fraction than other samples.

AC susceptibility of 200 °C annealed $FeTe_{1-x}S_x$:$O_2$ system samples is shown in Figure 7(b). Both the real and imaginary parts of AC susceptibility clearly exhibit the bulk superconductivity at around 9 K. All samples have nearly same $T_c$ values but with different volume fractions. The trend is same as seen for DC susceptibility in Figure 7(a) i.e., volume fraction is maximum for x = 0.30 and least for x =0.10. Now to check the impact of post $O_2$ annealing, we intercompare the results of Figure 7(b) with that of Figure (4 & 5), i.e. without $O_2$ annealing. It is clear that the volume fraction of $FeTe_{0.75}S_{0.25}$ sample is increased after $O_2$ annealing. Also there seems to be an increase in $T_c$ as well, although marginally of around 1 K only. This clearly indicates that volume fraction of weakly superconducting $FeTe_{0.75}S_{0.25}$ sample is improved after $O_2$ annealing. The exact role of $O_2$ annealing is not yet understood, except that there is slight decrease in volume suggesting towards the chemical pressure [18,19]. May it be that some S ions are substituted by O at 200 $^o$C or may remain at interstitial sites? In any of the case, post $O_2$ annealing improves superconductivity of $FeTe_{1-x}S_x$. In fact in a very recent article superconductivity in $FeTe_{1-x}S_x$ has been induced by alcohol as well [26]. The exact role of low temperature $O_2$ annealing or alcohol etc. is not yet understood. In the present article we report that to improve the superconductivity of $FeTe_{1-x}S_x$, one need not to necessary expose the samples to humidity for several days or anneal in $O_2$ for 100s hours, but the same could be achieved by just 12 hours annealing in $O_2$ at 200 $^o$C.

## Conclusion

We obtained the superconductivity in FeTe system by S substitution at Te site. The S substitution induced the superconductivity and suppressed the structural phase transition in the parent compound. The S substitution creates a positive chemical pressure in FeTe parent compound, which is responsible for the appearance of superconductivity in S substituted samples. The superconductivity of $FeTe_{1-x}S_x$ is further improved by post $O_2$ annealing at 200 $^o$C. Interestingly, dissimilar to several day humidity exposures [18] or long hours (100's) annealing; half a day low temperature (200 $^o$C) $O_2$ post annealing has improved the superconductivity of $FeTe_{1-x}S_x$ quite significantly and results are comparable to previous reports [18,19].

## Acknowledgement

Anand Pal and Arpita Vajpayee are thankful to CSIR for providing the financial support during his research. Authors acknowledge the support and encouragement from Director of the laboratory Prof. R.C. Budhani.

References

1. Y. Kamihara, T. Watanabe, M. Hirano and H. Hosono, J. Am. Chem. Soc. 130 (2008) 3296.
2. H. Takahashi, K. Igawa, K. Arii, Y. Kamihara, M. Hirano and H. Hosono, Nature 453 (2008) 376.
3. Z. A. Ren, W. Lu, J. Yang, W. Yi, X. L. Shen, Z. C. Li, G. C. Che, X. L. Dong, L. L. Sun, F. Zhou and Z. X. Zhao, Chin. Phys. Lett. 25 (2008) 2215.
4. M. Rotter, M. Tegel and D. Johrendt, Phys. Rev. Lett. 101 (2008)107006.
5. X. C. Wang, Q. Liu, Y. Lv, W. Gao, L. X. Yang, R. C. Yu, F. Y. Li, and C. Jin, Solid. State. Commun. 148 (2008) 538.
6. F. C. Hsu, J. Y. Luo, K. W. Yeh, T. K. Chen, T. W. Huang, P. M. Wu, Y. C. Lee, Y. L. Huang, Y. Y. Chu, D. C. Yan and M. K. Wu, Proc. Natl. Acad. Sci. U.S.A. 105 (2008) 14262.
7. Yoshikazu Mizuguchi, Fumiaki Tomioka, Shunsuke Tsuda and Takahide Yamaguchi, Appl. Phys. Lett. 93 (2008) 152505.
8. K. W. Yeh, T. W. Huang, Y. L. Huang, T. K. Chen, F. C. Hsu, P. M. Wu, Y. C. Lee, Y. Y. Chu, C. L. Chen, J. Y. Luo, D. C. Yan and M. K. Wu, Euro. Phys. Lett. 84 (2008) 37002.
9. M. H. Fang, H. M. Pham, B. Qian, T. J. Liu, E. K. Vehstedt, Y. Liu, L. Spinu and Z. Q. Mao, Phy. Rev. B 78 (2008) 224503.
10. Y. Mizuguchi, F. Tomioka, S. Tsuda, T. Yamaguchi and Y. Takano, J. Phys. Soc. Jpn. 78 (2009) 074712.
11. A. Subedi, L. Zhang, D. J. Singh and M. H. Du, Phys. Rev. B 78 (2008) 134514.
12. W. Bao, Y. Qiu, Q. Huang, M. A. Green, P. Zajdel, M. R. Fitzsimmons, M. Zhernenkov, M. Fang, B. Qian, E. K. Vehstedt, J. Yang, H. M. Pham, L. Spinu, and Z. Q. Mao, Phys. Rev. Lett. 102 (2009) 247001.
13. Shiliang Li, Clarina de la Cruz, Q. Huang, Y. Chen, J. W. Lynn, Jiangping Hu, Yi-Lin Huang, Fong-Chi Hsu, Kuo-Wei Yeh, Maw-Kuen Wu, and Pengcheng Dai1, Phys. Rev. B 79 (2009) 054503.
14. G. F. Chen, Z. G. Chen, J. Dong, W. Z. Hu, G. Li, X. D. Zhang, P. Zheng, J. L. Luo, N. L. Wang, Phys. Rev. B 79 (2009) 140509(R).

15. Y. Mizuguchi, F. Tomioka, S. Tsuda, T. Yamaguchi and Y. Takano, Physica C 469 (2009) 1027.
16. Hironari Okada, Hiroyuki Takahashi, Yoshikazu Mizuguchi, Yoshihiko Takano and Hiroki Takashi, J. Phys. Soc. Jpn. 78 (2009) 083709.
17. Yoshikazu Mizuguchi, Fumiaki Tomioka, Shunsuke Tsuda, Takahide Yamaguchi and Yoshihiko Takano, Appl. Phys. Lett. 94 (2009) 012503.
18. Y. Mizuguchi, K. Deguchi[1], S. Tsuda[1], T. Yamaguchi[1] and Y. Takano, Phys. Rev. B 81 (2010) 214510.
19. Y. Mizuguchi, K. Deguchi, S. Tsuda, T. Yamaguchi and Y. Takano, Euro. Phys. Lett. 90 (2010) 57002.
20. M. Nikolo and R. B. Goldfarb, Phys. Rev. B 39 (1989) 6615.
21. S. L. Shinde, J. Morrill, D. Goland, D. A. Chance and T. McGuire, Phys. Rev. B 41 (1990) 8838.
22. K. H. Muller, Physica C 168 (1990) 585.
23. V.P.S. Awana, R.S. Meena, Anand Pal, Arpita Vajpayee, K.V. R. Rao and H. Kishan, arXiv: 1003.0273v2 (2010).
24. G. Bonsignore, A. Agliolo Gallitto, M. Li Vigni, J. L. Luo, G. F. Chen, N. L. Wang and D. V. Shovkun, cond-mat arXiv:1005.3965v1 (2010).
25. Xiaolin Wang, Shaban Reza Ghorbani, Germanas Peleckis and Shixue Dou, Advanced Materials 21 (2009) 236.
26. Keita Deguchi, Yoshikazu Mizuguchi, Toshinori Ozaki, Shunsuke Tsuda, Takahide Yamaguchi, Yoshihiko Takano, arXiv:1008.0666 (2010).

**Figure Captions:**

Figure 1(a): The observed (red dots), calculated (solid line) and differences x-ray diffraction (bottom solid line) profiles at 300 K for FeTe$_{1-x}$S$_x$; x=0.0, 0.10, 0.25 & 0.30.

Figure 1(b): Variation of lattice parameters and cell volume with sulphur (S) content.

Figure 2: Temperature dependence of the resistance $R(T)$ for FeTe and FeTe$_{0.75}$S$_{0.25}$ samples measured in 0 kOe field, inset shows the heat capacity variation with temperature C$_p$(T) in zero-field, 7 and 14 Tesla field of FeTe sample.

Figure 3: Variation of thermoelectric power with temperature $S(T)$ for FeTe$_{0.75}$S$_{0.25}$ sample.

Figure 4: Plot of real (*M'*) and imaginary (*M"*) components of AC susceptibility as a function of temperature at the frequencies 33, 333, 999, 3333, 6666 & 9999 Hz for fixed amplitude of 0.5 Oe of FeTe$_{0.75}$S$_{0.25}$ sample.

Figure 5: Variation of real (*M'*) and imaginary (*M"*) components of the AC susceptibility versus temperature, measured in the FeTe$_{0.75}$S$_{0.25}$ sample at the AC field amplitudes 0.1, 0.5, 1, 2 & 4 at a fixed frequency 333 Hz and inset shows the enlarge view of *M"* versus *T* at 0.1 Oe for the same sample.

Figure 6: Temperature dependence of the resistance $R(T)H$ of FeTe$_{0.75}$S$_{0.25}$ sample measured in fields up to 110 kOe and inset shows the $H_{c2}$ vs *T* plots derived from $R(T)H$ plots using Ginzburg landau (GL) equation.

Figure 7(a): Temperature dependence of the DC magnetization of FeTe$_{1-x}$S:O$_2$ samples.

Figure 7(b): Temperature dependence of real and imaginary part of AC susceptibility measured under applied field of 10 Oe for FeTe$_{1-x}$S:O$_2$ samples.

Table1: The lattice constants, volume of unit cells, 'z' coordinates of Te/S and Rietveld refinement parameters for $FeTe_{1-x}S_x$ (x= 0.0 - 0.30)

| S.No | Sample | $a$ (Å) | $c$ (Å) | $V$ (Å$^3$) | Z(Te/S) | $R_p$(%) | $R_{wp}$(%) | $\chi^2$ |
|---|---|---|---|---|---|---|---|---|
| 1. | x = 0.0 | 3.821(9) | 6.285(5) | 91.765(23) | 0.2822(35) | 3.85 | 5.47 | 2.33 |
| 2. | x = 0.05 | 3.813(4) | 6.257(6) | 90.988(19) | 0.2830(27) | 3.47 | 4.84 | 2.09 |
| 3. | x = 0.10 | 3.811(9) | 6.239(3) | 90.617(13) | 0.2838(33) | 4.10 | 5.21 | 2.36 |
| 4. | x = 0.15 | 3.809(1) | 6.237(6) | 90.488(26) | 0.2843(56) | 3.10 | 4.01 | 1.72 |
| 5. | x = 0.20 | 3.808(7) | 6.229(1) | 90.317(14) | 0.2853(43) | 4.01 | 5.11 | 2.67 |
| 6. | x = 0.25 | 3.806(9) | 6.226(4) | 90.189(19) | 0.2864(29) | 3.47 | 4.69 | 2.41 |
| 7. | x = 0.30 | 3.805(7) | 6.224(2) | 90.103(20) | 0.2878(49) | 4.22 | 5.43 | 3.34 |

Figure 1(a):

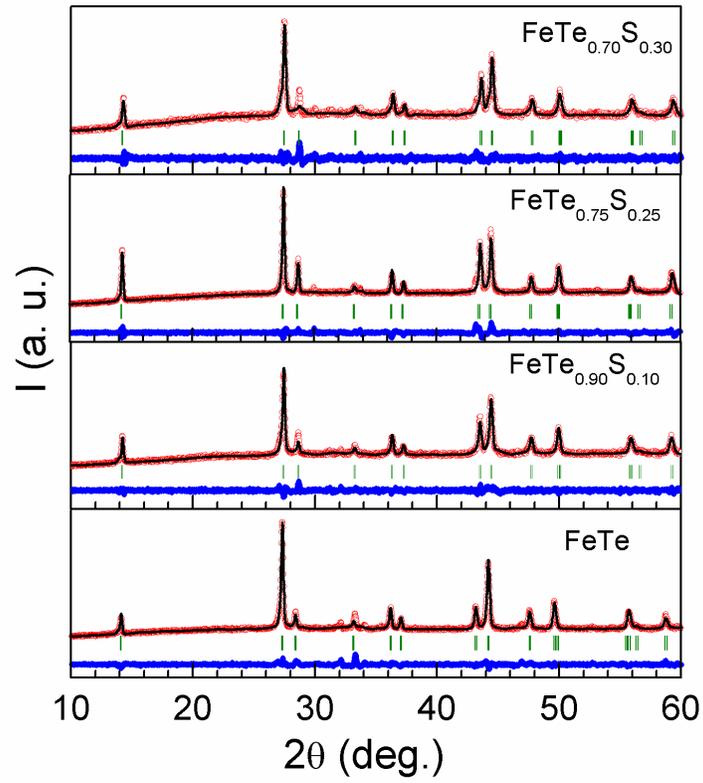

Figure 1(b):

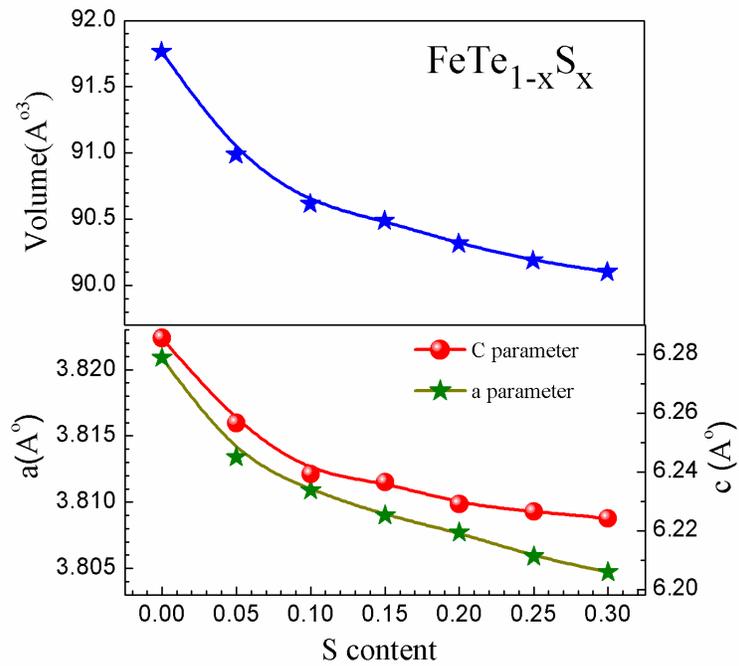

Figure 2:

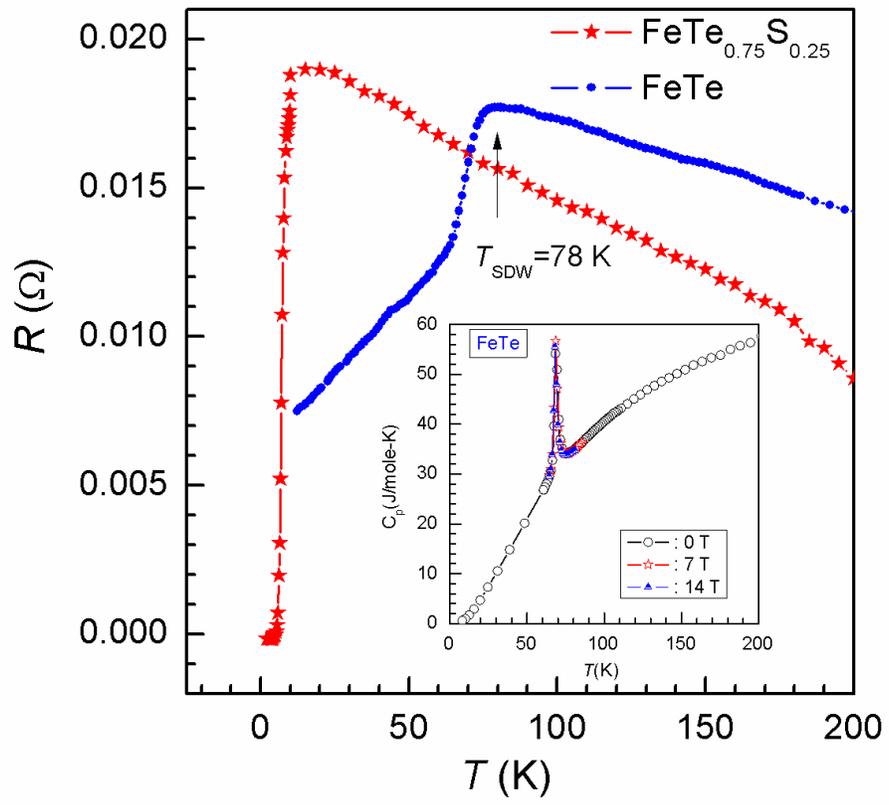

Figure 3:

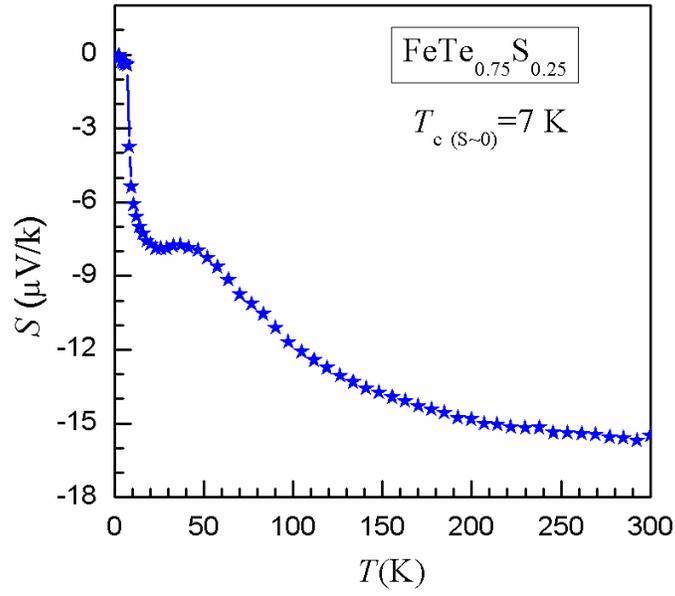

Figure 4:

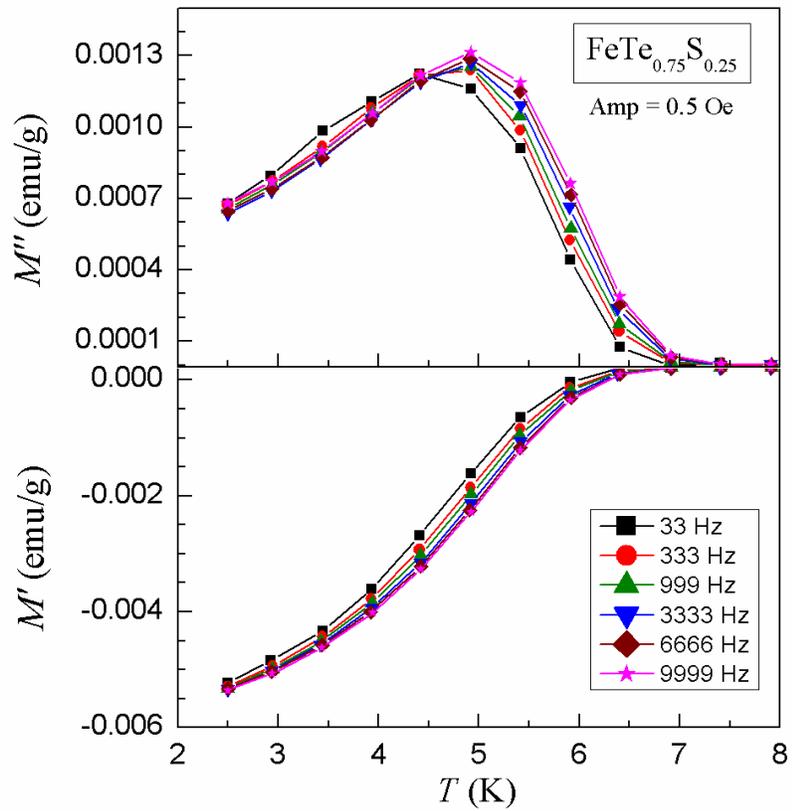

Figure 5:

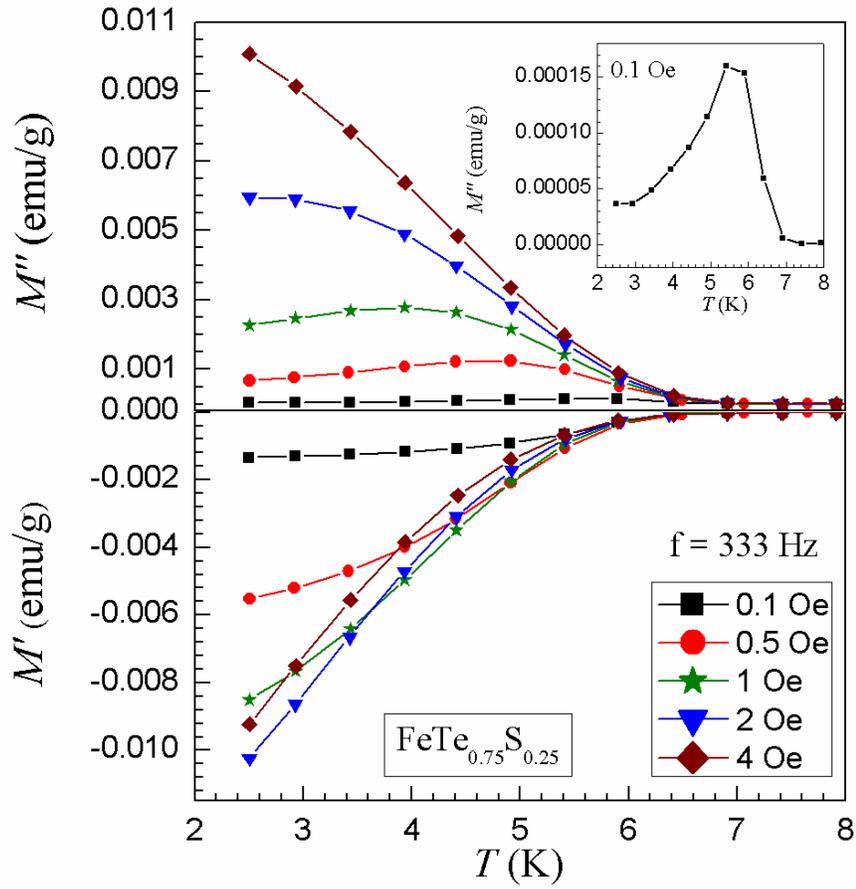

Figure 6:

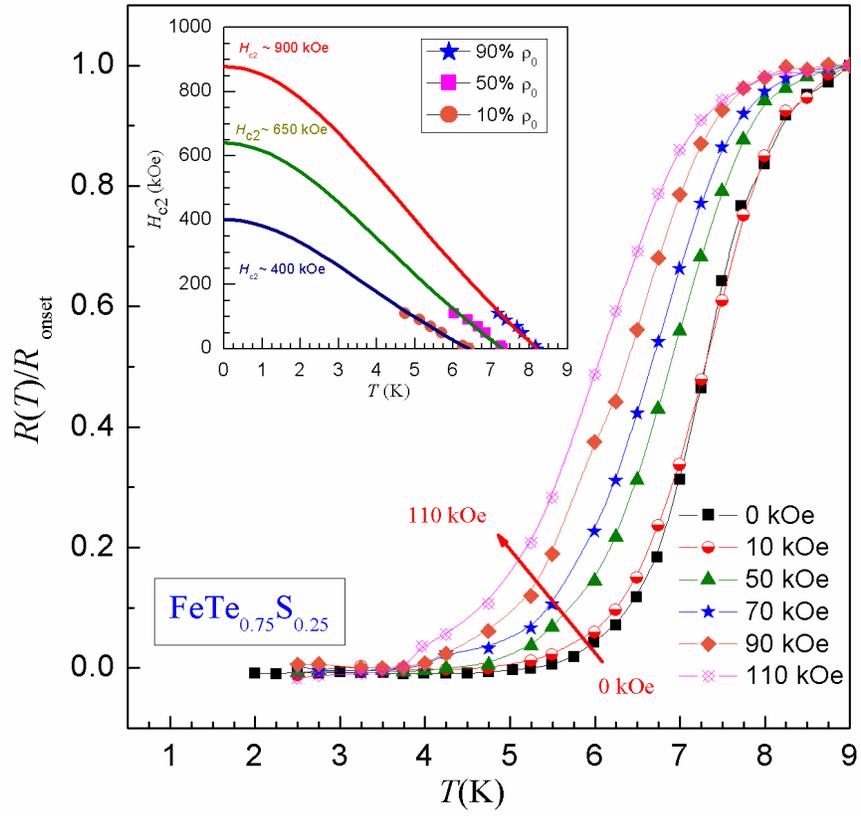

Figure 7 (a):

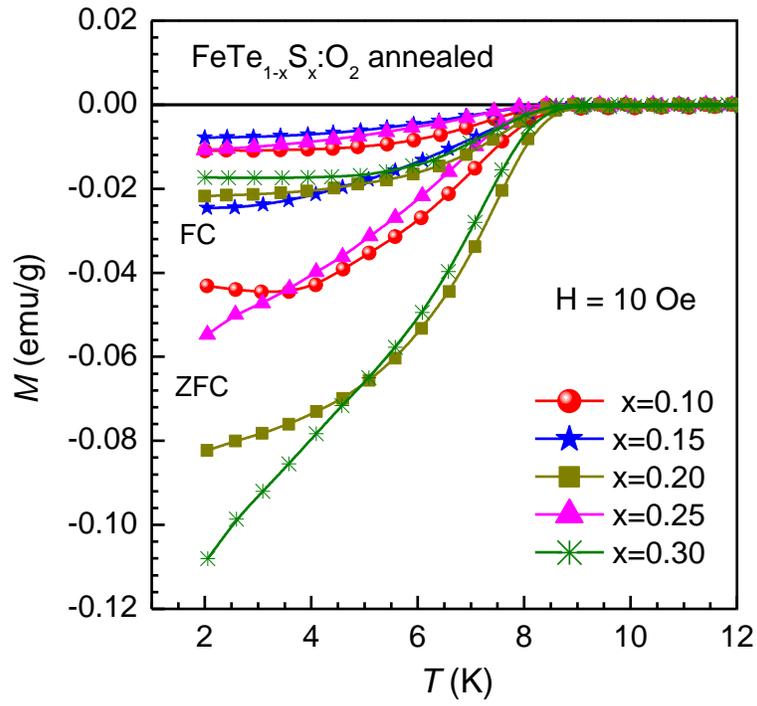

Figure 7 (b):

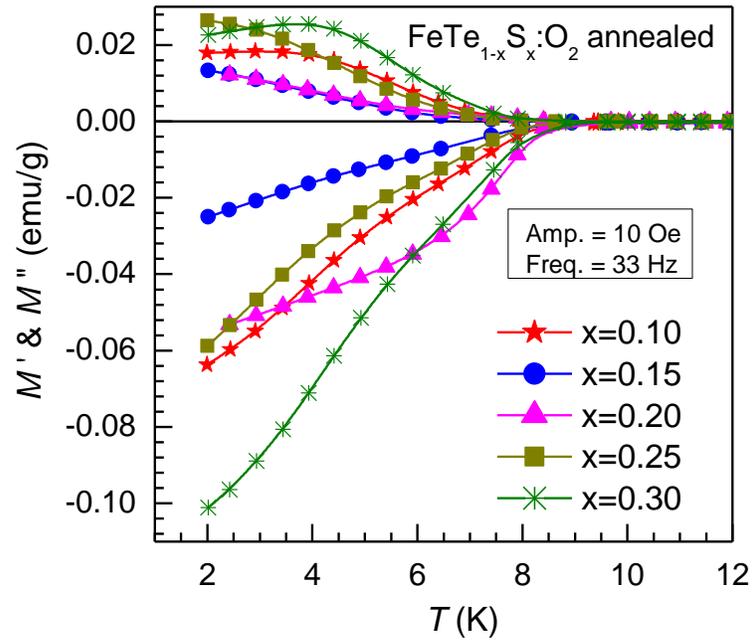